\documentclass[showpacs,floatfix,superscriptaddress,showpacs,twocolumn,amssymb,amsfonts,prb,aps]{revtex4}
\usepackage{longtable,graphicx,epsfig,dcolumn}

\begin{document}
\bibliographystyle{revtex}
\title
{Anomalous layering at the liquid Sn surface}

\author{Oleg~G.~Shpyrko}
\affiliation{Department of Physics, Harvard University, Cambridge
Massachusetts 02138 (USA)}

\author{Alexei~Yu.~Grigoriev}
\affiliation{Division of Engineering and Applied Sciences, Harvard
University, Cambridge Massachusetts  02138 (USA)}

\author{Christoph~Steimer}
\affiliation{Division of Engineering and Applied Sciences, Harvard
University, Cambridge Massachusetts  02138 (USA)}

\author{Peter~S.~Pershan}
\affiliation{Department of Physics, Harvard University, Cambridge
Massachusetts 02138 (USA)}\affiliation{Division of Engineering and
Applied Sciences, Harvard University, Cambridge Massachusetts  02138
(USA)}

\author{Binhua~Lin}
\affiliation{The Center for Advanced Radiation Sources, University
of Chicago, Chicago, Illinois 60637 (USA)}

\author{Mati~Meron}
\affiliation{The Center for Advanced Radiation Sources, University
of Chicago, Chicago, Illinois 60637 (USA)}

\author{Tim~Graber}
\affiliation{The Center for Advanced Radiation Sources, University
of Chicago, Chicago, Illinois 60637 (USA)}

\author{Jeff~Gebhardt}
\affiliation{The Center for Advanced Radiation Sources, University
of Chicago, Chicago, Illinois 60637 (USA)}

\author{Ben~Ocko}
\affiliation{Department of Physics, Brookhaven National Lab, Upton
New York 11973 (USA)}

\author{Moshe~Deutsch}
\affiliation{Department of Physics, Bar-Ilan University, Ramat-Gan
52900 (Israel)}

\date{\today}

\begin{abstract}

\def\baselinestretch{1}

 X-ray reflectivity measurements on the free surface of
liquid Sn are presented. They exhibit the high-angle peak,
indicative of surface-induced layering, also found for other pure
liquid metals (Hg, Ga, and In). However, a low-angle shoulder, not
hitherto observed for any pure liquid metal, is also found,
indicating the presence of a high-density surface layer.
Fluorescence and resonant reflectivity measurements rule out the
assignment of this layer to surface segregation of impurities. The
reflectivity is modeled well by a 10$\%$ contraction of the spacing
between the first and second atomic surface layers, relative to that
of subsequent layers. Possible reasons for this are discussed.

\end{abstract}


\maketitle

\section{Introduction}

Rice and co-workers\cite{Rice74} predicted that the atoms at the
free surface of a liquid metal should be stratified to a depth of a
few atomic diameters. This layering phenomenon was experimentally
confirmed two decades later by x-ray reflectivity measurements for
three high-surface-tension metals: Hg,\cite{Magnussen95} Ga
\cite{Regan95}, and In,\cite{Tostmann99} and the low-surface-tension
metal K.\cite{Shpyrko03} The signature of layering in the x-ray
reflectivity curve is the appearance of a quasi-Bragg peak at a wave
vector transfer $q_z = 2 \pi / d$, where $d$ is the atomic spacing
between the layers. The peak arises from constructive interference
of waves diffracted by the ordered surface layers. These
measurements are complicated by several technical issues, in
particular the need to use UHV conditions to preserve surface
cleanliness for highly reactive liquid metals surfaces.

Both Ga and In were found to exhibit a simple layering structure
described in detail by Regan \emph{et al.}\cite{Regan95} comprising
equal-density, periodically spaced atomic layers. Liquid
Hg\cite{Magnussen95,Dimasi98} shows a more complicated surface
structure. However, unlike Ga and In, the high vapor pressure of Hg
did not allow its study under UHV conditions, and the possibility
that the more complicated structure originates in chemical
interactions with foreign atoms at the surface cannot be
definitively ruled out. For Ga, in particular, the decay length of
the layering into the bulk was found to be slightly larger than that
expected, and found, for other liquid metals. This was tentatively
assigned to the enhanced Ga-Ga pairing tendency,\cite{Regan95}
reflected also in the appearance of a small shoulder on the first,
nearest-neighbor, peak of the bulk radial distribution
function,\cite{Dicicco94,Gong93} and in the ordering of Ga at the
liquid-solid interface.\cite{vanderVeen97} A similarly strong
pairing tendency has been reported also for Sn, the subject of the
present study.\cite{Jovic76, DiCicco96, Itami03}

The motivation of the present study was to investigate whether
atomic layering exists in liquid Sn  and, if it does, to find out
whether or not the layering follows the classic behavior of Ga and
In or do new effects appear, e.g. due to the strong pairing. The
relatively low melting temperature, $T_m = 232~^{\circ}$C and
vanishingly low vapor pressure at $T_m$ render such measurements
possible. As shown below, the measured x-ray reflectivity of the Sn
surface indeed exhibits an anomalous feature that is not present for
either Ga or In. Additional experiments were carried out to rule out
the possibility that this anomalous feature is not intrinsic, but
caused by chemical impurities at the surface. The origin of this
structure is identified as a $\sim 10$\% reduction in the first to
second interlayer distance.

\section{Experiment}

The UHV chamber and the preparation procedures of the liquid Sn
sample have been described previously.\cite{Tostmann99, Huber02}
Prior to placing an Sn sample inside the Mo sample pan, the surface
of the pan was sputtered clean. In the sputtering process the
sample's surface is bombarded by Ar$^{+}$ ions, which break up and
sputter away the oxidized material covering the surface. Ingots of
solid Sn with purity of 99.9999$\%$ were placed in the Mo sample pan
inside a UHV chamber evacuated to $10^{-9}$ Torr and were then
melted by a Boralectric heating element mounted underneath the pan.
After the molten Sn filled the pan, forming approximately
10-mm-thick liquid sample, it was cooled to the solid phase,
followed by a 24-h bakeout process. During such bakeout the walls of
the chamber and its components are gradually heated up to
$100-200^{\circ}$C and then cooled down to room temperature.
Following this procedure the sample was melted again and the
macroscopic native oxide, as well as any possible contaminations, at
the surface were removed by a mechanical scraping of the liquid
surface with Mo foil wiper. After this, the residual microscopic
surface oxide layer was removed by further sputtering by the
Ar$^{+}$ ion beam for several hours. If there are impurities in the
bulk with surface energies that are lower than that of Sn the Gibbs
adsorption rule implies that they would segregate at the surface. We
will demonstrate below that even though Sn has a relatively high
surface tension impurities are not present at the surface.

\section{X-ray reflectivity measurements}

\begin{figure}[tbp]
\epsfig{file=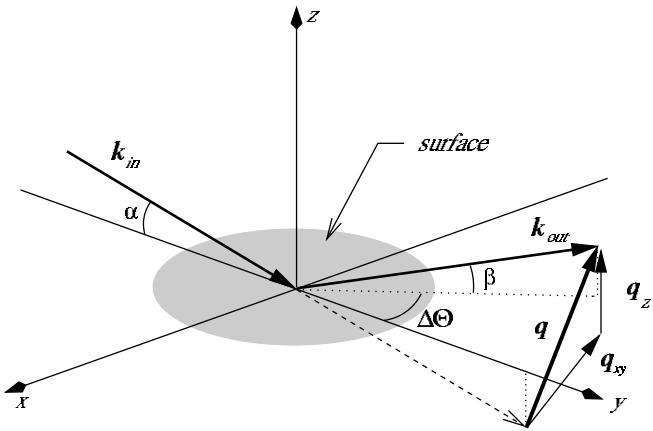, angle=0, width=1.0\columnwidth}
\caption{\label{fig:K_kinema}Kinematics of the x-ray measurement.
$k_{in}$\ and $k_{out}$ are the wave vectors of the incident and
detected x-rays, respectively.}
\end{figure}

The measurements were carried out at the liquid surface spectrometer
facility of the ChemMat~CARS beamline at the Advanced Photon Source,
Argonne National Laboratory, Argonne, IL. The kinematics of x-ray
measurements described in this work are illustrated in Fig.
\ref{fig:K_kinema}. X rays with a wave vector $k_{in}=2\pi/\lambda$,
where $\lambda=0.729$~{\AA}  is the x-ray wavelength, are incident
on the horizontal liquid surface at an angle $\alpha$. The detector
selects, in general, a ray with an outgoing wave vector $k_{out}$.
The reflectivity $R(q_z)$ is the intensity ratio of these two rays,
when the specular conditions $\alpha=\beta$\ and $\Delta\Theta=0$
are fulfilled. In this case the surface-normal momentum transfer is
$q_z=(2\pi/\lambda)(\sin \alpha+\sin \beta)=(4\pi/\lambda) \sin
\alpha$. The detector resolution due to a finite acceptance angle
was defined by two pairs of slits mounted on the detector arm and
during all specular reflectivity measurements was fixed at
$3.9~mrad$ vertically and $2.6~mrad$ horizontally. During the
measurements the temperature of the sample has been maintained at
$240~^\circ~C$, which is just above bulk melting temperature of Sn,
$232~^\circ~C$.

\begin{figure}[tbp]
\epsfig{file=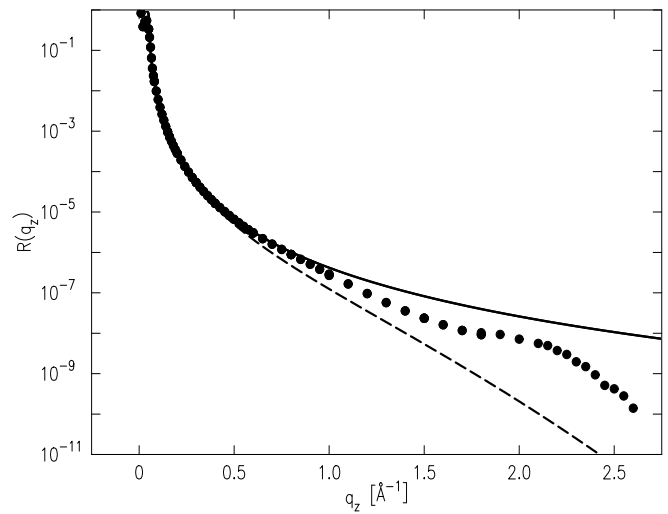, angle=0, width=1.0\columnwidth}
\caption{\label{fig:Snref} The measured x-ray specular reflectivity
(points) of the surface of liquid Sn. The Fresnel reflectivity
(solid line) of an ideally flat and abrupt surface and the
reflectivity of an ideal surface roughened by thermally excited
capillary waves (dashed line) are also shown.}
\end{figure}

The x-ray specular reflectivity shown with circles in
Fig.~\ref{fig:Snref} is the difference between the specular signal
recorded with an Oxford scintillation detector at $\alpha=\beta,
\Delta\Theta=0$ and the off-specular background signal recorded at
the same $q_z$, but at $\Delta \Theta=  \pm 0.1^\circ$. The data are
then normalized to the measured incident intensity. This background
subtraction procedure is particularly important because the bulk
scattering function peaks at approximately the same $q_z$ as the
quasi-Bragg surface layering peak, since both peaks correspond
roughly to the interatomic distance. As a result the background is
particularly strong at the $q_z$ values corresponding to the
layering peak. Since the measured $\Delta \Theta \neq 0$ intensity
includes contributions from the capillary-wave-induced diffuse
surface scattering, all of the theoretical simulations discussed
below include a similar background subtraction procedure. The
Fresnel reflectivity curve $R_F(q_z)$,  due to an ideally flat and
abrupt surface, is shown as a solid line in Fig.~\ref{fig:Snref}.
The dashed line is $R_F(q_z)$ modified by the theoretically
predicted thermal capillary wave contributions, which depend only on
the known values of the resolution function, surface tension and
temperature. The quasi-Bragg layering peak and the deviation of the
measured $R(q_z)$ from the capillary-wave-modified $R_F(q_z)$ are
unambiguous proof of the existence of local structure at the
surface. The quasi-Bragg peak at $q_z \approx 2.2$~{\AA}$^{-1}$
corresponds to an atomic layering of close-packed spheres with the
$d\approx2.8$~{\AA} spacing of the atomic diameter of Sn. An
additional new feature, a subtle but significant shoulder at $q_z
\approx 0.9$~{\AA}$^{-1}$ is not discernible in this figure and is
revealed only upon removal of the effects of the capillary waves
from the measured Fresnel-normalized reflectivity, as we show below.

\section{Surface Structure Factor}

\begin{figure}[tbp]
\epsfig{file=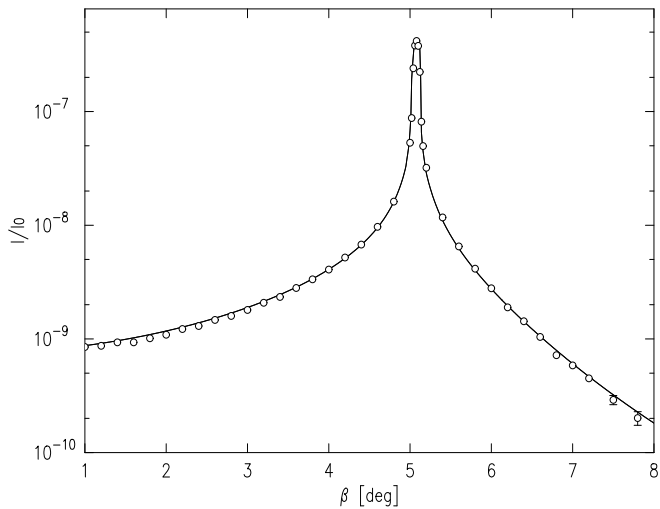, angle=0, width=1.0\columnwidth }
\caption{\label{fig:Sndiff} Diffuse scattering off the liquid Sn
surface measured at a fixed grazing angle of incidence $\alpha=$
5.05$^{\circ}$ (open circles). The peak corresponds to the
specular condition at $q_z=0.48$~{\AA}$^{-1}$. The line is the
capillary wave theory predictions for surface tension
$\gamma=560$~mN/m.}
\end{figure}

To obtain a quantitative measure of the intrinsic surface structure
factor the effects of thermal capillary excitations must be
deconvolved from the reflectivity curve shown in
Fig.~\ref{fig:Snref}. As mentioned above, the method for doing this
has been described in detail in earlier papers on the surface
layering in Ga,\cite{Regan95} In,\cite{Tostmann99} K
\cite{Shpyrko03}, and water.\cite{Shpyrko04} The principal result of
this analysis is that the measured reflectivity can be expressed as
\begin{equation}
R(q_z) = R_F(q_z) \cdot  | \Phi (q_z) | ^2 \cdot CW(q_z)
\label{eq:structure1}
\end{equation}
where $\Phi (q_z)$ is the intrinsic structure factor of the surface
and CW($q_z$) accounts for the effects of the thermally excited
capillary waves on $R(q_z)$. The surface scattering cross section
yielding the CW($q_z$) term depends on $T$, $q_z$, the surface
tension and the geometric parameters defining  the reflectometer's
resolution function. It is understood well enough to allow us to
fully account analytically for the effects of capillary waves on the
measured $R(q_z)$. This is demonstrated in Fig.~\ref{fig:Sndiff}
where we plot the background-subtracted and
incident-intensity-normalized intensity scattered by the surface
capillary waves measured as a function of $\beta$ for a fixed
incidence angle $\alpha$ in the reflection plane
($\Delta\Theta=0^\circ$). The agreement between the measured values
(points) and the theoretical cross section (line) is very good over
more than three decades in intensity.

For this scan the incident beam is 100 $\mu$m high and the detector
is 0.5~mm high and 1~mm wide. Integration of the intensity in this
figure over the $\beta$ range spanned by the resolution function of
the reflectivity measurements shown in Fig.~\ref{fig:Snref} yields a
value identical with that obtained in the reflectivity measurements
at $q_z=0.48$~{\AA}$^{-1}$. This further strengthens our claim that
the effects of the capillary-wave scattering on the reflectivity are
well understood and can be separated out confidently from the
measured $R(q_z)$.

The iterative procedure by which the theoretical line in
Fig.~\ref{fig:Sndiff} is calculated is the following. We first
obtain a measure of the structure factor from a best fit of
Eq.~\ref{eq:structure1} to the reflectivity (Fig.~\ref{fig:Snref})
using values for CW($q_z$) that are calculated from published values
for the surface tension. This form of the surface structure factor
is then used to fit measured diffuse scattering curves similar to
that shown in Fig.~\ref{fig:Sndiff}, with the surface tension as the
only adjustable parameter. When needed, the cycle can be repeated
using now the new surface tension value. In practice, however, a
single cycle was enough, since the best fit value for the surface
tension, $\gamma=560$~mN/m, was in good agreement with the published
values. Thus, we assert that the form of the diffuse scattering that
gives rise to the value of $CW(q_z)$ is fully determined, and the
only unknown quantity in Eq.~\ref{eq:structure1} is the surface
structure factor. This quantity is the Fourier transform of the
surface-normal derivative of the local electron density $\rho(z)$:
\cite{Braslau88}
\begin{figure}[tbp]
\epsfig{file=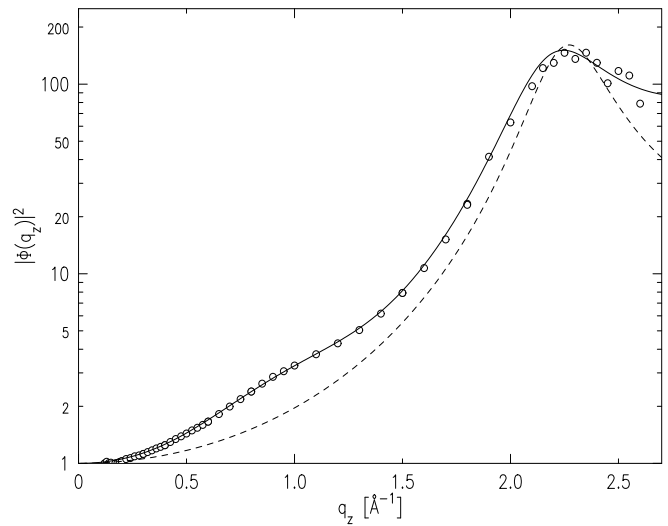, angle=0, width=1.0\columnwidth}
\caption{\label{fig:Sncw} The structure factor (squared) of the Sn
surface as derived from the measured reflectivity $| \Phi (q_z) |
^2 =R(q_z)/[R_F(q_z)CW(q_z)]$ (open circles). The dashed line is
the theoretically expected $| \Phi (q_z) | ^2$ of a simple layered
density profile as found previously for Ga and In. The solid line
is a fit to a model, discussed in the text, where the distance
between the first and second layers is reduced by 10\% relative to
that of the subsequent layers.}
\end{figure}
\begin{equation}
\Phi (q_z) = \frac{1}{\rho_{\infty}} \int dz\frac{d \langle
\rho(z) \rangle }{dz} \exp(\imath q_z z) \label{eq:structure2}
\end{equation}
where $\rho_{\infty}$ is the electron density of the bulk and
$\langle ...\rangle$\ denotes averaging over the surface-parallel
($x,y$) directions.

The ratio $R(q_z)/[R_F(q_z)CW(q_z)]=| \Phi (q_z) | ^2 $ derived from
the measured reflectivity, is plotted in Fig.~\ref{fig:Sncw} (open
circles). The dashed line is the theoretical reflectivity calculated
using the surface structure factor obtained from the simple atomic
layering model that successfully described the intrinsic surface
structure factor of the pure liquid Ga and In. As can be seen from
the plot, this model does not describe the measured values well. In
particular, it does not exhibit any feature corresponding to the
weak but distinct shoulder observed in the measured values at $q_z
\approx 0.9$~{\AA}$^{-1}$. As this shoulder's position is
incommensurate with the large layering peak at 2.2~{\AA}$^{-1}$, it
must indicate the existence of a second length scale, additional to
the periodicity of the surface-induced layering. It is highly
unlikely that this second length scale  $2\pi/0.9 = 6.96$~{\AA}  is
associated with a \emph{periodic} structure, in the same way that
the 2.2~{\AA}$^{-1}$ peak is associated with the surface-induced
layering, since such two-periodicity structure would be very
difficult to rationalize physically. Similarly, the low-$q_z$
shoulder cannot be assigned to just a single-periodicity layered
structure with a top layer which is denser than the subsequent
layers. While such a structure will yield a subsidiary peak at low
$q_z$, this peak would be commensurate with the high-$q_z$ one, i.e.
appear at $q_z=0.5 \times 2.2 \approx 1.1$~{\AA}$^{-1}$\ rather than
the observed $\sim 0.9$~{\AA}$^{-1}$. Additionally, our attempts to
fit the data using alternative density models - such as keeping the
periodicity intact while modifying electron densities of the top
several layers - did not produce a satisfactory fit to the
experimental reflectivity data. Thus, we conclude that an
appropriate model needs to include a second, noncommensurate and
nonperiodic length scale. The solid line shown in
Fig.~\ref{fig:Sncw} is a fit of the measured data by a model
constructed along these lines. As can be seen, a good agreement is
achieved with the measured values. We now proceed to discuss the
details of this model.

\begin{figure}[tbp]
\epsfig{file=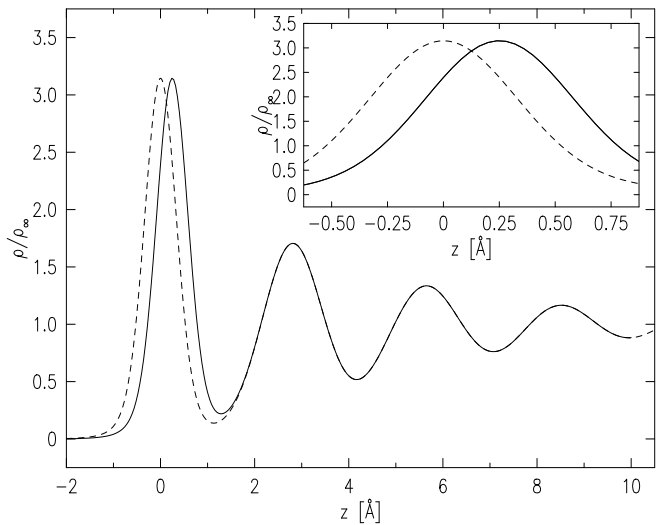, angle=0, width=1.0\columnwidth}
\caption{\label{fig:Snden} Models for the intrinsic surface-normal
electron density profiles of Sn for a simple, equally spaced
layering model (dashed line), and for a model including a
contraction of the spacing between the first and second layers by
10\% (solid line). The models yield the reflectivities shown by the
same lines in Fig.~\ref{fig:Snref}. The inset is a blow up of the
surface layer region.}
\end{figure}

The simple density model that has been used to represent the
reflectivity of Ga\cite{Regan95} and In\cite{Tostmann99} consists of
a convolution of an intrinsic density profile and a
Debye-Waller-like factor, which accounts for the smearing of this
profile by the thermally excited capillary waves. The intrinsic
profile is modeled by a semi-infinite series of equally spaced
Gaussians, each representing a single atomic layer. The Gaussians
have equal integrated areas (i.e. equal areal electron density) but
their widths increase with depth below the surface. This, in turn,
results in a gradual decrease with depth of the individual peaks and
valleys of the density profile, and an eventual evolution of the
density towards the constant average density of the bulk liquid. The
decay length of the surface layering is typically of the order of
just a few atomic spacings.\cite{Regan95, Tostmann99} The density
profile of this model is shown in a dashed line in
Fig.~\ref{fig:Snden} and yields the dashed line in
Fig.~\ref{fig:Sncw}. The solid line that runs through the measured
values in Fig. \ref{fig:Sncw} is calculated from a slightly modified
model, shown in a solid line in Fig. \ref{fig:Snden}. The only
difference between this model and the original one is that the
distance between the first and second layers, $2.55$~{\AA}, is
smaller by $\sim 10${\%} than the $2.8$~{\AA} spacing of the
subsequent layers. The average density over the first two layers is
thus larger than that of the bulk, and a second, nonperiodic, length
scale is introduced. The good agreement of this minimally-modified
model with the measured data, demonstrated in in
Fig.~\ref{fig:Sncw}, strongly supports our interpretation of the
surface structure on liquid Sn.

\section{THE CASE FOR EXCLUSION OF SURFACE IMPURITIES }

Up to this point we have not considered the possibility that the
dense surface layer may be a layer of atoms of a different metal
adsorbed onto the Sn surface. This is precisely what happens, for
example, in the liquid binary alloys GaBi\cite{Tostmann00b, Huber02}
and Ga-Pb,\cite{Yang99}, where a (dense) monolayer of the
lower-surface-energy species (Bi,Pb) is found to Gibbs-adsorb at the
free surface of the alloy. We now show experimental evidence that
this is not the case here.

Consideration of the Gibbs rule\cite{Gibbs28} shows that relatively
low concentrations of three metals, Bi ($\gamma$= 378 mN/m, $\rho_e
= 2.49$~e/{\AA}$^{3}$), Pb ($\gamma$= 458~mN/m, $\rho_e =
2.63$~e/{\AA}$^{3}$) and Tl ($\gamma$= 464~mN/m, $\rho_e =
2.79$~e/{\AA}$^{3}$) could have produced surface electron densities
sufficient to cause the low-$q_z$ shoulder shown in Fig.
\ref{fig:Sncw}. In order to address this possibility we performed
two independent experiments, grazing incidence x-ray fluorescence
(GI-XRF) and resonant x-ray reflectivity, both of which are
sensitive to the presence of sub monolayer quantities of impurities
at the surface. None of the two measurements found evidence for
contamination of the liquid Sn surface by a foreign species.
Therefore, this supports the conclusion that the anomalous density
feature observed here is in fact an intrinsic property of the liquid
Sn surface.

\subsection{Grazing incidence fluorescence scattering}
\begin{figure}[tbp]
\epsfig{file=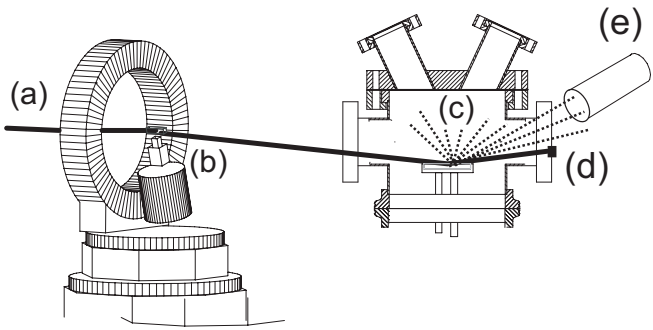, angle=0, width=1.0\columnwidth}
\caption{\label{fig:gid} Experimental setup used in the GI-XRF
measurements: the synchrotron beam (a) is Bragg-reflected from a
monochromator (b) which directs it downwards at the liquid Sn sample
located inside a UHV chamber (c). The specularly reflected beam
(solid line) is blocked by a beam stop on the exit window (d) to
minimize the background. The fluorescence from the sample's surface
(dotted lines) is detected off specularly by an energy-dispersive
detector (e) located as close to the chamber as possible to maximize
its solid angle of acceptance.}
\end{figure}

The experimental setup used in these measurements is shown in Fig.
\ref{fig:gid}.
The sample was illuminated by x rays of wavelength
$\lambda=0.42$~{\AA} (E=29.5~keV), incident at a grazing angle of
$\alpha=0.03^{\circ}$. Since this is approximately one-third of the
critical angle of Sn at this energy, the refracted wave is
evanescent. Its penetration depth below the surface is given by
\cite{Tolan99} $\tau = 1/$Im$[(4\pi/\lambda)\sqrt{\alpha ^2 -
\alpha_{crit} ^2-\imath\lambda\mu/2\pi} ]$, where the critical angle
for Sn is $\alpha_{crit}=$~0.09$^{\circ}$ and the linear absorption
coefficient is $\mu=3.26 \times 10^{-6}$~{\AA}$^{-1}$. For these
values the incident beam probes only the uppermost $\tau =
30$~{\AA}\ of the sample. Due to the high surface tension, the
surface of the liquid Sn sample is curved, which has to be taken
into account as well: for an incident beam height H and a convex
sample with a radius of curvature $r \approx 10$~m the angle of
incidence $\alpha$ varies over the illuminated area by $\sim
H/(\alpha r)$. In these measurements $H$ was set to 5~$\mu$m, so
that for a nominal angle of incidence $\alpha_0 =0.03^{\circ}$ the
local incidence angle varies by slightly less than $\pm
0.03^{\circ}$. Over this range, $0^{\circ}<\alpha<0.06^{\circ}$, the
penetration length $\tau$ varies from 21~{\AA}\ to 28~{\AA}. Thus,
for our nominal incident angle of $\alpha\approx 0.03^{\circ}$ any
detected fluorescence signal comes mostly from the top 8-10 atomic
layers. Depending on the signal-to-noise ratio and other factors,
such as the relative fluorescence yield, with this geometry it is
possible to detect trace amounts of selected materials to
sub-monolayer accuracies, as we show below. Note that at our
incident energy, 29.5~keV, which is well above the Sn K edge, the K
lines of Sn are all excited. However, the K edges of Bi, Pb and Tl
are all $>29.5$~keV, so that only the L lines of these elements are
excited.

Fluorescent x-ray emission from the illuminated portion of the
sample was detected by an energy dispersive intrinsic Ge detector
(area $\sim 9$~mm$^2$) that was mounted about 30~cm away from the
center of the sample, 15 cm above it, and displaced azimuthally by
about 30$^{\circ}$ from the incidence plane. The experimental setup
for this geometry is shown in Fig.~\ref{fig:gid}. The signal was
recorded using a multichannel analyzer. A lead shield was mounted on
the output window as shown in the Fig.~\ref{fig:gid} to eliminate
scattering from the specularly reflected beam into the detector by
the exit window or outside air. Unfortunately, there was no simple
way to shield the detector from scattering or fluorescence
originating in the entrance window of the UHV chamber. In order to
account for this background scattering, a differential measurement
has been performed. The signal detected with the sample displaced 3
mm below the incident beam was subtracted from the signal that was
measured for the beam striking the sample.
\begin{figure}[tbp] \epsfig{file=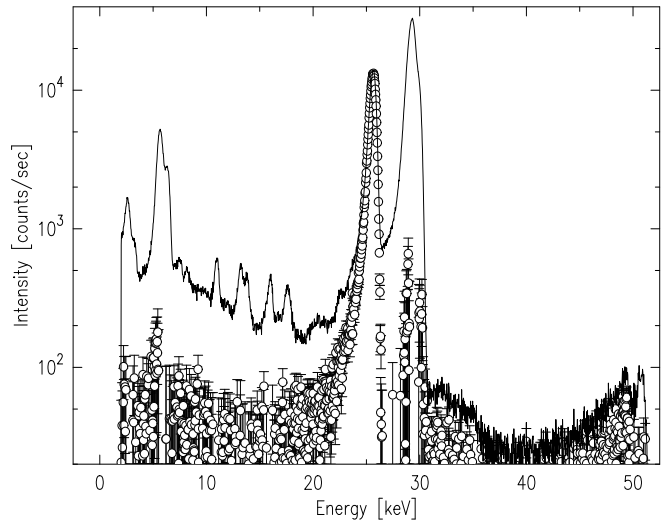,
angle=0, width=1.0\columnwidth} \caption{\label{fig:Snflu} Raw
(line) and background-subtracted (open circles) measured
fluorescence from the surface of liquid Sn. The sharp lines are
due to the K emissions lines\cite{Bearden67} of Fe (E=6.4~keV), Nb
(E=16.6~keV), Mo (E=17.5~keV) and Sn (E=25.2~keV).}
\end{figure}
The fluorescence data are shown in Fig.~\ref{fig:Snflu}. The solid
line is the as-measured raw fluorescence spectrum. It consists of
fluorescence from both the sample and the entrance window, elastic
scattering from the incident beam (the strongest peak at $\sim
29.15$~keV), and Sn fluorescence (at 25.2~keV). There are a number
of lower-energy peaks originating in the UHV chamber's components.
The open circles show the spectrum obtained after subtracting from
the raw spectrum the background spectrum recorded when the sample is
moved out of the beam. The only remaining prominent peak is the Sn
fluorescence observed at 25.2~keV. No other characteristic lines are
discernible above the noise level. The ratio of the Sn
K-fluorescence to the noise level in the 10-15~keV energy region,
where the L-fluorescence lines of the Pb, Bi or Tl are expected, is
$\sim$200. Factoring in the ratio of the K fluorescence yield of Sn
(0.84) to the L-fluorescence yields of Pb, Bi or Tl ($\sim$0.4)
indicates that the presence of as little as 10$\%$ of a full
monolayer of these metals (out of the $\sim$10 atomic layers of Sn
illuminated by the evanescent wave) should be detectable in this
experiment. Since an impurity coverage of the surface required to
generate the $\sim 0.9$~{\AA}$^{-1}$\ reflectivity peak is
significantly higher than that, the absence of an impurity signal in
Fig.~\ref{fig:Snflu} rules out with a high degree of confidence the
possibility that a Gibbs-adsorbed layer of impurities is the
originator of the $0.9$-{\AA}$^{-1}$ feature.

\subsection{Resonant x-ray reflectivity}
The effective electron density of an x-ray-scattering atom is
proportional to the scattering form factor $f(q)+f'(q,E)$. Here
$f(q) \rightarrow Z$ for $q \rightarrow 0$ and $Z$ is the atomic
number. When the x-ray energy is tuned through an absorption edge of
a scattering atom, the magnitude of the real part of $f'$ undergoes
a sharp decrease, producing a change in the scattering power of that
specific atom.\cite{Sparks94, Dimasi00} Thus, by measuring the
scattering on and off edge it is possible to isolate the scattering
due to the specific atom. We have used this method, called resonant
(or anomalous) scattering, in the reflectivity mode to probe the
liquid surface of Sn for the presence of foreign atoms. This is done
by comparing the x-ray reflectivities of the liquid surface measured
with the x-ray energy tuned on and off the K edge of Sn (29.20 keV).
If the low-$q_z$ feature at $q_z \approx 0.9$~{\AA}$^{-1}$ is due a
thin surface layer of foreign atoms, the effective electron density
of Sn at the edge will change, while that of the foreign atoms will
not. Thus, the density contrast will change and so will the
prominence of the low-$q_z$ feature. If, however, the low-$q_z$
shoulder is due to an intrinsic Sn structure, the electron density
\emph{difference} between high-density surface layer and the bulk
will remain unchanged and so will the low-$q_z$ shoulder.

\begin{figure}[tbp]
\epsfig{file=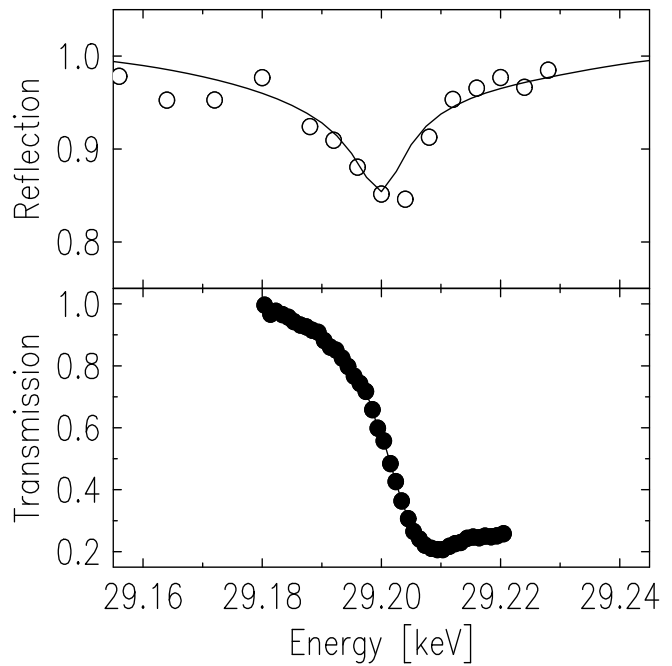, angle=0, width=1.0 \columnwidth}
\caption{\label{fig:Snescan} (Top) Relative energy variation of
the reflectivity off the liquid Sn surface at a fixed
$q_z=0.3$~{\AA}$^{-1}$. The line is the theoretical prediction of
the change in the reflectivity due to the variation of the
dispersion correction $f'(E)$ of the scattering factor near the
edge, as calculated by the IFEFFIT\cite{ifeffit} program. The open
circles are the measured values. (Bottom) Measured energy
variation of the intensity transmitted through an Sn foil
(points). The K edge of Sn at E=29.20~keV shows up clearly in both
measurements.}
\end{figure}

The 29.20~keV edge of Sn was identified by a transmission
measurement through a Sn foil, shown in Fig. \ref{fig:Snescan}(b),
as well as by the reflectivity from the liquid Sn surface at a fixed
$q_z=0.3$~{\AA}$^{-1}$, shown in Fig. \ref{fig:Snescan}(a). Since
the surface structure factor $\Phi (q_z)$ depends on the electron
density contrast between surface and bulk, the only energy
dependence of $R(q_z)$ away from both structural peaks ( at
$q_z=0.9$~{\AA}$^{-1}$\ and $q_z=2.3$~{\AA}$^{-1}$) is due to the
energy variation of the critical wave vector $q_c$. This leads to a
$(Z+f'(E))^2$ energy dependence of the reflectivity at a fixed
$q_z$. The minimum value of $f'(E)$, estimated from Fig.
\ref{fig:Snescan}(a), is consistent with the theoretically predicted
value calculated by the IFEFFIT\cite{ifeffit} program, taking the
smearing due to the lifetime widths of the K and L levels into
consideration. The observed change in reflectivity at the minimum
relative to that 50~eV away is measured to be about 15\%, which
corresponds, in turn, to a 7\% reduction in the effective electron
density of Sn.

\begin{figure}[tbp]
\epsfig{file=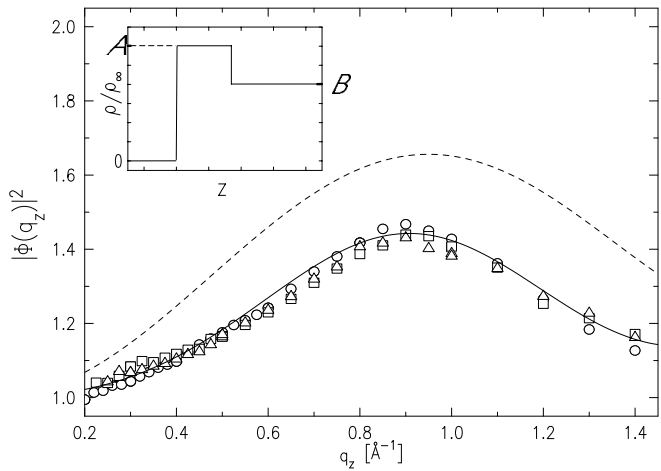, angle=0, width=1.0\columnwidth}
\caption{\label{fig:Sncwnorm} The structure factor (squared) of the
Sn surface as derived from the measured reflectivity, at
(E=29.20~keV, triangles), above (E=29.22~keV, circles) and below
(E=17~keV, squares) the K-edge. The solid line is a fit to the
density model featuring reduced interatomic spacing for the top
layer by approximately 10$\%$. The dashed line is calculated for a
model with a high-density surface monolayer of an atomic species
other than Sn. The inset shows the density profile of such a layer.
For a discussion see the text.}
\end{figure}

In Fig.~\ref{fig:Sncwnorm} we show the low-$q_z$ normalized
reflectivity data $R(q_z)/[R_F(q_z)CW(q_z)]=| \Phi (q_z) | ^2 $,
measured at the K edge of Sn (29.20 keV), just above the edge (29.22
keV) and far below the edge (17 keV). The fact that the three data
sets coincide proves unambiguously  that the $0.9$-{\AA}$^{-1}$
reflectivity feature does not arise from contrast between Sn and a
different element. A more quantitative analysis can be done in two
ways. First, we assume a simple box model for the
surface,\cite{Tolan99} shown in the inset to
Fig.~\ref{fig:Sncwnorm}. In this model, the surface layer's electron
density A is larger than that of the bulk, B, yielding $| \Phi (q_z)
| ^2 \approx (2A/B-1)^2$\ at the peak. To obtain the measured peak
of $| \Phi (q_z) | ^2 = 1.4$, we have to assume $A/B=1.1$. We note
that only a handful of elements would be consistent with a model
where this 10\% increase in density originates in a surface layer of
impurity atoms. Eliminating the elements which have a higher surface
tension than that of Sn ($\gamma$=560 mN/m), which will not
Gibbs-segregate at the surface, the only reasonably likely remaining
candidates are Bi, Pb and Tl. If we assume that the surface layer is
a monolayer of any one of these elements, the effect of the
resonance is that the effective electron density of the Sn would be
reduced by about $7\%$, as mentioned above. The measured
$R(q_z)/[R_F(q_z)CW(q_z)]=| \Phi (q_z) | ^2 $ would therefore
increase at the Sn K-edge to $(2\times 1.1/0.93 - 1)^2=1.86$. The
fact that this does not occur proves, again, that the increased
density at the surface is an intrinsic surface effect, due entirely
to a structure comprising Sn atoms only.

A second quantitative argument is the full modeling of the
reflectivity, rather than considering only the maximum of the
$0.9$-{\AA}$^{-1}$ peak. The solid line in Fig.~\ref{fig:Sncwnorm}
is the same fit shown in Fig.~\ref{fig:Snden} to the model where the
density of each atomic layer is kept at the Sn density value, while
the spacing between the first and second atomic layers is reduced
from the 2.80~{\AA} of subsequent layers to 2.55~{\AA}. The
agreement of all measured reflectivities, in particular that
measured at the edge (triangles), with this model is excellent. The
dashed line is the reflectivity calculated assuming an additional
surface monolayer of a different species having a density higher
than that of Sn. As can be seen, for the on-edge measurement, this
line is expected to lie considerably higher than off edge. It
clearly does not agree with the on-edge measured data (triangles).

The results of the two measurements presented in this section rule
out the possibility that the dense surface layer found in the
reflectivity measurements is due to surface segregation of impurity
atoms. Thus, they strongly supports our claim that the high-density
layer present at the surface is an intrinsic property of Sn itself.

\section{CONCLUSION}

We have found that the surface of liquid Sn exhibits atomic
layering, similar to that found in other metallic systems. However,
the surface structure factor exhibits an additional low-$q_z$
shoulder, which has not been observed for any of the pure metals
studied to date. The possibility that this structure is due to a
layer of surface-adsorbed contaminants has been addressed by two
experiments, both of which confirm that the observed reflectivity
feature is an intrinsic property of liquid Sn. The strongest
argument comes from the resonant x-ray reflectivity measurement
which finds no detectable difference between reflectivities measured
on and off the Sn K edge, while model calculations predict a
significant change for a layer of foreign surface atoms. Our
analysis shows that the anomalous feature at low $q_z$ is consistent
with a modified atomic layering density model where the spacing
between the first and second atomic layers is reduced by 10$\%$
relative to the subsequent ones, thereby increasing the average
density at the surface.

The physical reasons for the existence of this denser layer at the
surface of liquid Sn are not clear, since the studies of other
liquid metals and alloys exhibit no evidence for such a feature. It
should be noted that the observed spacing reduction is similar to
the well-known surface relaxation phenomena found in many
crystalline monatomic metals. Surface relaxation typically manifests
itself in a reduction of the lattice spacing between the surface
atomic layer and first subsurface atomic layer commonly of the order
5$\%$ to 10$\%$  \cite{Kittel_book}, which is comparable to the
spacing reduction found in this study. Additionally, surface
relaxation is generally enhanced for the metals exhibiting a high
degree of covalent bonding\cite{Kadas96}, while Sn-Sn pairing is
well known to be present in the bulk phase \cite{Jovic76, DiCicco96,
Itami03} of liquid Sn. And since contrary to solids the surface
reconstruction mechanism is not possible at liquid surfaces due to
absence of outward-oriented dangling bonds at the surface layer,
surface relaxation remains as the primary mechanism of minimizing
the surface energy associated with a reduced number of nearest
neighbors at the surface.

Another argument to consider comes from Sn having a rather
complicated crystalline structure below melting point: the density
of the liquid Sn (6970~kg/m$^3$) is in between the densities of the
solid $\alpha$-Sn (5750~kg/m$^3$), also known as "gray tin" and
forming a cubic atomic structure, and the solid $\beta$-Sn
(7310~kg/m$^3$), known as "white tin" and forming a tetragonal
atomic structure. It is possible that while the higher-density
$\beta$~phase is no longer stable in the bulk beyond the melting
point, it prevails in some form at the surface, where the packing
restrictions are relaxed due to a smaller number of nearest
neighbors.

\section{ACKNOWLEDGMENTS}
We thank Matt Newville, Larry Sorensen, and John Rehr for helpful
discussions on IFEFFIT use. This work has been supported by U.S.
Department of Energy Grant No. DE-FG02-88-ER45379, National Science
Foundation Grant No. DMR-01-12494, and the U.S.--Israel Binational
Science Foundation, Jerusalem. ChemMatCARS Sector 15 is principally
supported by the National Science Foundation/Department of Energy
under Grant No. CHE0087817. The Advanced Photon Source is supported
by the U.S. Department of Energy, Basic Energy Sciences, Office of
Science, under Contract No. W-31-109-Eng-38.


\bibliographystyle{unsrt}

\begin{thebibliography}{10}

\bibitem{Rice74}
S.A. Rice, D.~Guidotti, and H.L. Lemberg.
\newblock  {\em Aspects of the study of
  surfaces}, (Wiley, Chichester, UK, 1974) vol.~27, pp. 543-633.

\bibitem{Magnussen95}
O.~M. Magnussen, B.~M. Ocko, M.~J. Regan, K.~Penanen, P.~S.
Pershan, and  M.~Deutsch.
\newblock  Phys. Rev. Lett. \textbf{74}, 4444 (1995).

\bibitem{Regan95}
M.~J.~Regan, E.~H.~Kawamoto, S.~Lee, P.~S.~Pershan, N.~Maskil,
M.~Deutsch, O.~M.~Magnussen, B.~M.~Ocko, and L.~E.~Berman
\newblock  Phys. Rev. Lett. \textbf{75}, 2498 (1995).

\bibitem{Tostmann99}
H.~Tostmann, E.~DiMasi, P.~S. Pershan, B.~M. Ocko, O.~G. Shpyrko,
and M.~Deutsch.
\newblock  Phys. Rev. B \textbf{59}, 783 (1999).

\bibitem{Shpyrko03}
O.~Shpyrko, P.~Huber, A.~Grigoriev, P.~Pershan, B.~Ocko,
H.~Tostmann, and  M.~Deutsch \newblock Phys. Rev. B
  \textbf{67}, 115405 (2003).

\bibitem{Dimasi98}
E.~Dimasi, H.~Tostmann, B.~M.~Ocko, P.~S.~Pershan and M.~Deutsch
\newblock Phys. Rev. B \textbf{58}, 13419 (1998).

\bibitem{Dicicco94}
A.~Di~Cicco and A.~Filliponi
\newblock Europhys. Lett. \textbf{27}, 407 (1994).

\bibitem{Gong93}
X.~S.~Gong et al.
\newblock Europhys. Lett. \textbf{21}, 469 (1994).

\bibitem{vanderVeen97}
W.~J.~Huisman, J.~F.~Peters, M.~J.~Zwanenburg, S.~A.~deVries,
T.~E.~Derry, D.~Abernathy, J.~F.~van~der~Veen
\newblock Nature \textbf{390}, 379 (1997).

\bibitem{Jovic76}
D.~Jovic and I.~Padureanu.
\newblock  Journal of Physics C \textbf{9}, 1135 (1976).

\bibitem{DiCicco96}
A.~DiCicco.
\newblock Phys. Rev. B \textbf{53}, 6174 (1996).

\bibitem{Itami03}
T.~Itami, S.~Munejiri, T.~Masaki, H.~Aoki, Y.~Ishii, T.~Kamiyama,
Y.~Senda,  F.~Shimojo, and K.~Hoshino.
\newblock Phys. Rev. B \textbf{67}, 064201 (2003).

\bibitem{Huber02}
P.~Huber, O.~G.~Shpyrko, P.~S.~Pershan, B.~M.~Ocko, E.~DiMasi, and
M.~Deutsch
\newblock Phys. Rev. Lett. \textbf{89}, 035502 (2002).

\bibitem{Shpyrko04}
O.~G.~Shpyrko, M.~Fukuto, B.~M.~Ocko, P.~S.~Pershan, M.~Deutsch,
T.~Gog and I.~Kuzmenko.
\newblock Phys. Rev. B \textbf{69}, 245423, (2004).

\bibitem{Braslau88}
A.~Braslau, P.~S. Pershan, G.~Swislow, B.~M. Ocko, and
J.~Als-Nielsen.
\newblock  Phys. Rev. A \textbf{38}, 2457 (1988).

\bibitem{Tostmann00b}
H.~Tostmann, E.~DiMasi, O.~G. Shpyrko, P.~S. Pershan, B.~M. Ocko,
and M.~Deutsch.
\newblock Phys. Rev. Lett. \textbf{84}, 4385 (2000).

\bibitem{Yang99}
B.~Yang, D.~Li, Z.~Huang, and S.~A.~Rice
\newblock Phys. Rev. B \textbf{62}, 13111 (2000).

\bibitem{Gibbs28}
J.~W.~Gibbs, R.~G.~Van Name, W.~R.~Longley, and H.~A.~Bumstead,
\newblock {\em Collected works of J. Willard Gibbs}, (Longmans, Green and Co., New York, 1928).

\bibitem{Tolan99}
M.~Tolan.
\newblock {\em X-Ray Scattering from Soft-Matter Thin Films}, (Springer-Verlag New York, 1999) p. 148.

\bibitem{Bearden67}
J.~A.~Bearden.
\newblock Rev. Mod. Phys., \textbf{39}, 78-124 (1967).

\bibitem{Sparks94}
C.~J. Sparks and K.~Fischer.
\newblock {\em Resonant Anomalous X-ray Scattering: Theory and Applications}, \newblock(North-Holland, Amsterdam, 1994).


\bibitem{Dimasi00}
E.~DiMasi, H.~Tostmann, O.~G.~Shpyrko, M.~Deutsch, P.~S.~Pershan,
and B.~M.~Ocko.
\newblock Journal of Physics: Condensed Matter \textbf{12}, 209 (2000).

\bibitem{ifeffit}
Matt Newville
\newblock {IFEFFIT: Interactive XAFS Analysis,
  http://cars9.uchicago.edu/ifeffit/}, 1997.

\bibitem{Kittel_book} C.~Kittel, \newblock {\em
Introduction to Solid State Physics}, Wiley, New York, 1996.

\bibitem{Kadas96}
K.~Kaddas, S.~Alvarez, E.~Ruiz, P.~Alemany, \newblock Phys. Rev. B
\textbf{53}, 4933 (1996).

\end{thebibliography}

\end{document}